\documentclass[prl,twocolumn,showpacs,preprintnumbers,amsmath,amssymb,notitlepage]{revtex4-1}
\usepackage{graphicx}
\usepackage{ulem}
\UseRawInputEncoding

\begin{document}

\title{ Remarks on Statistical mechanics of a moving system   }
\author{  Jinwu Ye  }
\affiliation{ $^{1}$ The School of Science, Great Bay University, Dongguan, Guangdong, 523000, China   \\
$^{2}$  Department of Physics and Astronomy, Mississippi State University, MS, 39762, USA     }
\date{\today }


\begin{abstract}
In the realm of statistical mechanics, it has been established that there is no distinction between the micro-canonical and canonical ensembles in the thermodynamic limit. However, this paradigm may alter when addressing statistical mechanics in the context of a moving sample with a velocity 
$ v $. Our investigation reveals significant disparities between the two ensembles when considering relativistic effects up to the order of  $ (v/c)^2 $.
While the temperature remains the same in the former, it experiences an increase in the latter.
If the system undergoes a finite-temperature phase transition, the critical temperature decreases in the co-moving frame of the latter ensemble. 
The implications of these findings on the thermodynamic zeroth to the third laws and the eigenstate thermalization hypothesis are analysed.
The potential for the experimental detection of these novel effects in  condensed matter systems are discussed.
\end{abstract}

\maketitle

{\sl 1. Introduction. }
The realms of thermodynamics and quantum statistical mechanics offer descriptions of the collective behaviors of large numbers of interacting particles from a phenomenological and microscopic perspective, respectively. The  temperature is the fundamental physical quantity to characterize these behaviors. 

Conventional textbooks \cite{book1,book2,kardarbook,Pbook} assert that, in the thermodynamic limit, there are no distinctions between the micro-canonical and canonical ensembles when portraying thermal disordered statistical behaviors. Furthermore, it has been established that the partition function of a quantum statistical mechanics system can be effectively mapped to thermal quantum field theory, articulated through a path integral formalism in imaginary time 
at a finite temperature\cite{scaling,QPTbook}. However, understanding how these statistical behaviors change when the system is in motion remains a fundamental open problem.

This study addresses this intriguing problem by employing a thermal quantum field theory approach. We directly apply a Lorentz transformation to the partition function to investigate the alterations in temperature induced by the system's motion. 
This framework  can be used to describe the shift of the thermal phase transition at $ T_c >0 $ in a moving body
if the system hosts such a phase transition in the static sample.
It may also be readily extended to Quantum Field Theory (QFT) in a curved space-time for the exploration of quantum black holes.

Examining the micro-canonical and canonical ensemble, we observe that the temperature of a moving body remains unchanged and increases respectively. 
In the case of a system exhibiting a finite temperature phase transition, such as a 3D XY transition in a static superfluid within a canonical ensemble, the critical temperature in the co-moving frame of a running superfluid decreases.
We delve into the implications of these findings on the thermodynamic zeroth to the third laws, including the generalization of the zeroth law to a moving body.
Possible intrinsic connections with the eigenstate thermalization hypothesis (ETH)  are explored. 
Differences and limitations between the experimental set-ups in  condensed matter systems and those in particle physics are analyzed.
Feasible experimental detections of these novel effects in condensed matter systems  such as a running superfluid are outlined.




\begin{figure}[tbhp]
\centering
\includegraphics[width=.9 \linewidth]{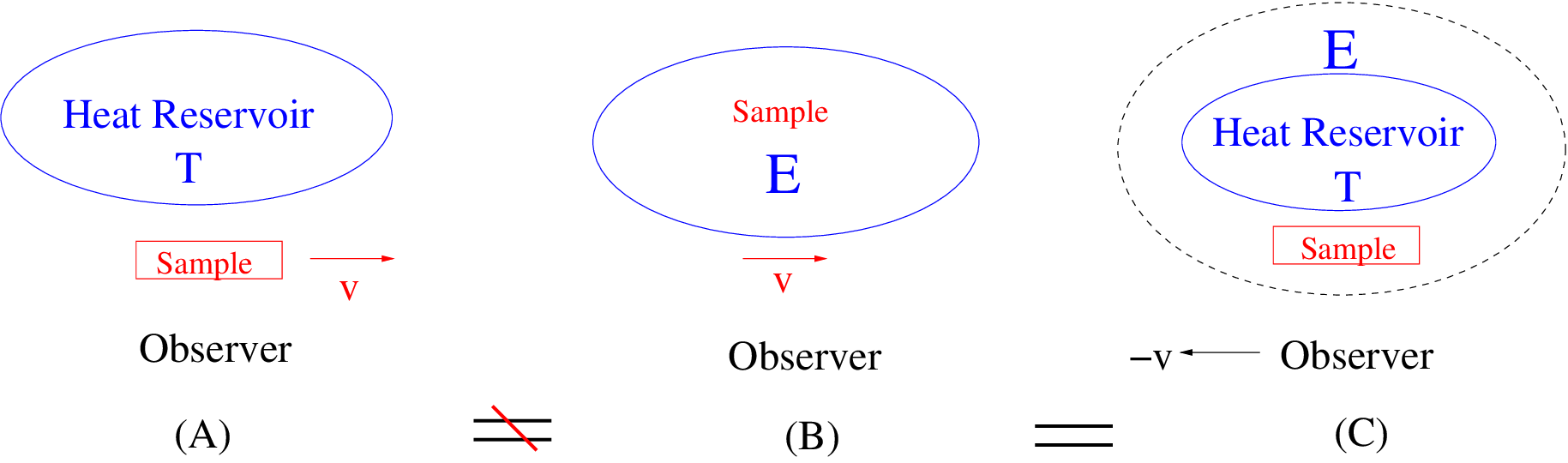}
\caption{  Canonical ensemble (A) versus the micro-canonical ensemble in (B) in the moving case. 
(B) is equivalent to (C) when one binds the reservoir and the system into an isolated system with the  total energy $ E $.
There is a heat exchange  between the system and the reservoir  in (A), but just an isolated system in (B).
Most condensed matter experiments belong to (A). All particle physics experiments belong to (B) or (C). }
\label{relativemoving}
\end{figure}

{\sl 2. Imaginary time  in the thermal QFT. }
 In a static sample relative to the reservoir in Fig.\ref{relativemoving},  the  partition function $  {\cal Z}[\beta, \mu]= Tr e^{-\beta H } $ in a canonical ensemble
  of any relativistic or non-relativistic quantum field theory  at any temperature $ T $  can be written in a path-integral with the Euclidean signature as:
\begin{equation}
 {\cal Z}= \int {\cal D} \phi e^{ - {\cal S} },~~~{\cal S}[ \beta ]= \int^{\beta}_0 d \tau \int d^d x {\cal L }[\phi, \partial_\mu \phi ]
\label{actionfiniteT}
\end{equation}
  where $ \beta=1/k_B T $ is the inverse temperature set-up by a reservoir. For simplicity, we only write the Lagrangian $  {\cal L } $   dependance on the complex bosons, 
  it can be easily generalized to include fermions and gauge fields. For the particle experiments, 
  $ {\cal L} $ is Lorentz invariant (LI) at  zero temperature $ \beta =\infty $.
  However, $ {\cal L} $ is not LI in any condensed matter or cold atom systems.
  The following derivation on the  law of $ T $  transformation    is general and universal,
  independent of any microscopic details such as the specific form of the Hamiltonian $ H $ or its Lagrangian $ {\cal L} $, 
  or if they are LI or not. At zero temperature $ \beta =\infty $,  one perform 
  $ SO(3,1) $ or $ SO(4) $ Lorentz transformation $ x^{\prime}= \Lambda x $  on  $ {\cal Z} $ between any two inertial frames 
   in the Lorentzian and  Euclidean signature respectively.

\begin{figure}[tbhp]
\centering
\includegraphics[width=.7 \linewidth]{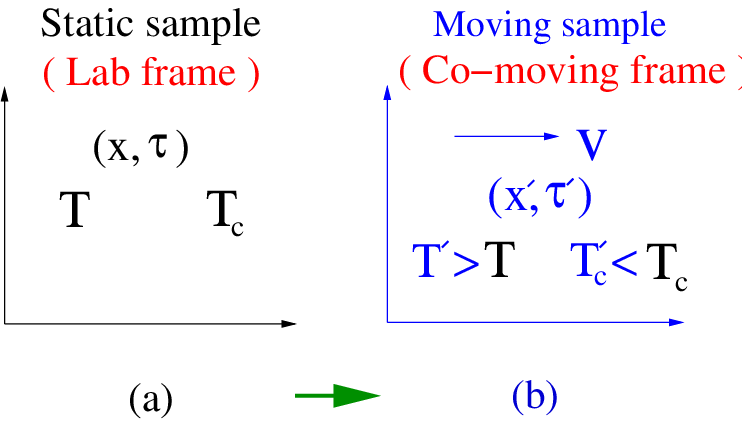}
\caption{ How does the temperature change is given by the $ SO(4) $ rotation from the static to the moving sample.
 (a) In a static sample relative to the reservoir,  $ (\tau, x ) $ is the space-time,
 its temperature is $ T $.
 (b) In a moving sample relative to the reservoir, $ (\tau^{\prime}, x^{\prime} ) $ is the space-time,
 its temperature is $ T^{\prime} $.  $ T^{\prime}/T > 1 $ is given by Eq.\ref{Tlaw}. 
 The critical temperature in the co-moving frame $ T^{\prime}_c/T_{c} < 1 $ is given by Eq.\ref{Tclaw}. }
\label{labcomoving}
\end{figure}

  At any finite temperature,  it is always more convenient to get to the  Euclidean signature by going to the imaginary time $ \tau= i t $.
  For simplicity, we take the Lorentz boost along the $ x- $ direction with the velocity $ v $:
\begin{align}
  x    & =\gamma ( x^{\prime}- i v \tau^{\prime} )    \nonumber   \\
  \tau & =\gamma ( \tau^{\prime}+ i \frac{v}{c^2} x^{\prime} )
\label{imagtauc}
\end{align}
 where $ \gamma=(1- (v/c)^2)^{-1/2} $ and $ (\tau, x ) $ and  $ (\tau^{\prime}, x^{\prime} ) $ are the space-time in the static sample and in a moving sample
 with the velocity $ v $  respectively ( Fig.\ref{labcomoving} ).

 Assuming the temperature in the static sample  $ (\tau, x ) $ is $ \beta $. What is the corresponding temperature
 in the moving sample  $ (\tau^{\prime}, x^{\prime} ) $  ?
 The quickest way to identify the temperature of a system is to identify the ( anti-)periodicity along the imaginary time direction
 for bosons or fermions respectively.  Just like the proper time or the proper mass ( the rest mass ) are LI, we expect the
 temperature  $ T^{\prime} $ determined in the co-moving frame are LI, so it can also be called the proper Temperature.

{\sl 3. Temperature transformation: }
 In Eq.\ref{imagtauc}, the two events in the imaginary time $ (\tau_1=0, x_1 ), (\tau_2= \beta,x_2 \neq x_1 ) $  corresponds to
\begin{align}
  0 & =\gamma (  \tau^{\prime}_1 + i \frac{v}{c^2} x^{\prime} )  \nonumber   \\
  \beta & =\gamma (  \tau^{\prime}_2 + i \frac{v}{c^2} x^{\prime} )
\label{imagtauxc}
\end{align}
  Due to assuming the same position $ x^{\prime} $ in the co-moving frame, one finds:
\begin{align}
 \beta= \gamma \Delta \tau^{\prime},~~~~~~\Delta \tau^{\prime}= \gamma^{-1} \beta
\label{imagtauxtc}
\end{align}
   which can also be called the proper temperature. It leads to
\begin{equation}
   \frac{T^{\prime}}{T} =  1/\sqrt{1 - (\frac{v}{c})^2 }  \sim 1 + \frac{1}{2} (v/c)^2+\cdots
\label{Tlaw}
\end{equation}
  So a moving-body gets hotter.   It has a clear physical meaning: when there is no relative motion between the body and the reservoir,
  the body's temperature is the same as that dictated by the reservoir. However, when there is a relative motion, then
  the body's temperature  is higher than that of the reservoir.
  In fact, we can take this new effect as the generalization of the thermodynamic zeroth law in the static case to the moving case.
  Namely, when the sample is moving relative to the reservoir, the sample must keep a higher temperature than its
  reservoir to keep the equilibrium between them, the temperature difference is given by the relative velocity.
  Because $ T=0  \to T^{\prime}=0 $, the ground state remains the ground state, so the third law remains apply. 
  In fact, as to be shown below Eq.\ref{ZpZSP}, the first to the third law remains to apply in any initial  frame.

  In the non-relativistic limit $ c \to \infty $, it still reduces to $ T^{\prime}=T $.
  When $ v/c \to 1^{-} $, $ T^{\prime} \to \infty $ approaches to infinite temperature.
  Of course, at sufficiently high temperatures, the QFT itself may break down.

  Due to assuming the same position $ x^{\prime} $ in the moving sample, one can also see the difference in the static sample:
$
 \Delta x= x_2-x_1=\gamma (-i v \Delta \tau^{\prime})=-i v \beta
$
  which hints the result Eq.\ref{Tlaw} is correct. Because one must identify the
  periodicity in the imaginary time direction in the moving sample at the same $ x^{\prime} $.
  It may correspond to two different positions in the static sample which do not cause any concern.


Note that the $ i $ in Eqs.\ref{imagtauc}-\ref{imagtauxc} is just due to the imaginary time $ \tau=i t $ in the Euclidean signature.
It shows that the temperature is entangled with the complex space. So both time and space need to be expanded to the two dimensional complex plane.
So the original real 4-dimensional space-time is extended to the 4-dimensional complex space-time which corresponds to 8-dimensional real space-time.
This is very similar to a real manifold is extended to a complex manifold which must be even-dimensional.
However, despite the imaginary time is just the temperature,  it is still not known if the complex space has any direct experimental detectable effects.


{\sl 4. The thermodynamics in a moving sample:  }
 In fact, (A) and (B) in Fig.\ref{relativemoving} are just two special cases.
 In a generic case,  the sample  and the observer are moving  with velocity $ \vec{v}_1 $ and $ \vec{v}_2 $ respectively.
 The law of $ T $ transformation Eq.\ref{Tlaw} is completely determined by the relative motion between the sample and the reservoir,
 independent of any observer. However, the Thermodynamics involves not only $ T $ which is an intensive quantity,
 but also the volume $ V $ ( in the co-moving frame ) which is an extensive quantity and does depend on an observer. Then one must specify the observer
 when discussing the Thermodynamics. In the following, we discuss case (A) first where $ \vec{v}_1=v, \vec{v}_2 = 0 $,
 then the case (B) where $ \vec{v}_1 = 0, \vec{v}_2 = v $, then the combined case of (A) and (B) where $ \vec{v}_1 \neq 0, \vec{v}_2 \neq 0 $.
 In the co-moving frame, the observer is moving together with the sample  $ \vec{v}_1 = \vec{v}_2 \neq 0 $,
 so it is a special case of the combination of (A) and (B).

{\sl (a) The case A: }

 In the case (A), the observer is static with respect to the reservoir, the corresponding partition function  is
\begin{align}
  {\cal Z}_{A}= {\cal Z}^{\prime}[ \gamma^{-1}_1 \beta,   \gamma^{-1}_{1} V ]
\label{ZpZVA}
\end{align}
 In the case (A), the relative motion between the sample and the reservoir is
 identical to that between  the sample and the observer,
 so the same factor $ \gamma^{-1}_1 $ multiplying the two arguments   $ \beta $ and $ V $ in Eq.\ref{ZpZVA} which can be viewed as a finite 
 size in the imaginary time and space respectively ( Fig.\ref{steady}b ).
 Note that despite the two arguments are universal, independent of any microscopic details of the Lagrangian $ {\cal L} $, the form
 of the partition function  $ {\cal Z}^{\prime} $ do depend on them, so do the two conjugate variables $ S $ and $ P $  listed in Eq.\ref{ZpZSP}.
 Only for a LI system at $ T=0 $ such as particle physics experiments, $ {\cal Z}^{\prime}={\cal Z}  $.
 For any condensed matter systems,  $ {\cal Z}^{\prime} \neq  {\cal Z}  $. But as stressed at very beginning, 
 How does the temperature change is independent of these microscopic details. 


{\sl (b) The case B: }

 In the case (B) which is equivalent to the case (C), there is no relative motion between the sample and the reservoir, so 
 setting $ v=0 $ in Eq.\ref{Tlaw} just reduces to
\begin{align}
  \frac{T^{\prime}}{T } =  1
\label{TT0}
\end{align}
 
  The corresponding partition function in the case (B) is:
\begin{align}
  {\cal Z}_{B}= {\cal Z}^{\prime}[\beta,  \gamma^{-1}_{2} V ]
\label{ZpZVB}
\end{align}
  where the system's volume $ V $  will be viewed the same as Eq.\ref{ZpZVA} in the case (A). 


{\sl (c) The combining case of (A) + (B) and the co-moving case: }

  Now we can also combine the case (A) with the case (B) to the most general case:
  If assuming the sample and the observer are moving parallel to each other: $\vec{v}_1 = v_1 \hat{x}, \vec{v}_2 = v_2 \hat{x} $,
  then Eq.\ref{ZpZVA} can be extended to:
\begin{equation}
 {\cal Z}_{AB}= {\cal Z}^{\prime}[ \gamma^{-1}_1 \beta,  \gamma^{-1}_{12} V ]
\label{ZpZV12AB}
\end{equation}
  where $  v_{12}= \frac{v_1-v_2}{1- v_1 v_2/c^2 } $ is the relative 4-velocity between the observer and the sample.
 
 In the case  $\vec{v}_1 $ and $ \vec{v}_2  $ are not parallel, then the two Lorentz transformation from
 the system to the reservoir given by the boost $ K_2 $ and the reservoir to the observer given by the boost $ K_1 $ may commute
 only upto a rotation in the plane spanned by the two vectors $\vec{v}_1 $ and $ \vec{v}_2  $, called Thomas procession.
 Then the rotation still reflects in the Doppler shift
 where it may be an classical analog of the non-Abelian Berry phase in topological phases \cite{kane} to some extent.
 Indeed, its effects have been observed in the atomic experiments. 
  However, such an ambiguity does not exist in the present case:   just from the physical picture in Fig.1,
  the order has been fixed: the temperature is fixed first, the Doppler shift second, so $ \vec{v}_1  $  first, $ \vec{v}_2 $ second.
  
 As mentioned below Eq.\ref{ZpZVA},  $ \beta $ and $ V $ in Eq.\ref{ZpZV12AB}  can be viewed as a finite 
 size in the imaginary time and the real space respectively. This geometric view leads to the unified  geometric interpretation in Fig.\ref{steady}b
 where $ \gamma_1 $ and $ \gamma_{12}  $ are determined by the relative motion between the reservoir and the sample/
 the sample and the observer respectively ( simplified as R-S and S-O respectively in the following ).
 So in general, they scale differently.  The difference comes from the  motion of the observer relative to the reservoir.

 Setting $ T=0 $, Eq.\ref{ZpZV12AB}  depends only on $ \gamma^{-1}_{12} $, so it
 recovers to the LI case where the trinity problem reduces to the bipartite problem.
 Setting $ v_2=0 $ recovers to the case (A), $ v_1=0 $ recovers to the case (B),
 $ v_1=v_2 $ leads to the co-moving frame which we will focus in the following section.

{\sl 5. The transformation of the critical temperature $ T_c $ in the co-moving frame:  }
 In the co-moving case $ v_1=v_2 $, Eq.\ref{ZpZV12AB} simplifies to:
\begin{equation}
 {\cal Z}_{Co}= {\cal Z}^{\prime}[ \gamma^{-1}_1 \beta, V ]= {\cal Z}[ \gamma^{-1}_1 \beta, V ]
\label{ZCo}
\end{equation}
where $\gamma_{12}=1 $ and  $ \gamma_1  $ is solely determined by the R-S.
This fact indicates attaching a thermometer to the moving body can still record the  change of the temperature 
of the moving body, but of course, not the Doppler shift or any other effects anymore. So one can focus on the temperature effects only.

Eq.\ref{ZCo} seems complementary to the case (B) in Eq.\ref{ZpZVB} which sees only Doppler shift, but no temperature effects. 
One new and salient feature  in the co-moving frame is 
that  $ {\cal Z}^{\prime}= {\cal Z} $ in Eq.\ref{ZCo}. 
We assume there is a singularity at $ {\cal Z}( \beta_{c} ) $ indicating a critical temperature $ T_{c} $
in the static sample in the thermodynamic limit $ V \to \infty $. Then we find the
corresponding critical temperature in the co-moving frame in the moving sample is:
\begin{equation}
   \frac{T^{\prime}_c}{T_{c}} =  \sqrt{1 - (\frac{v}{c})^2 }  \sim 1 - \frac{1}{2} (v/c)^2 +\cdots
\label{Tclaw}
\end{equation}
 which is just opposite to Eq.\ref{Tlaw}.
  It has a clear physical meaning: when there is no relative motion between the body and the reservoir,
  the phase transition happens at the reservoir's temperature $ T_{c} $.
  However, when there is a relative motion, as shown in Eq.\ref{Tlaw}
  the body's temperature  is higher than that of the reservoir,
  then the phase transition happens at the reservoir's temperature $ T^{\prime}_{c} < T_{c} $. 
  It indicates the critical temperature $ T_c $ drops. Its applications to 3d XY transition and 2d KT transition in 
 a running Helium4 superfluid will be presented in the experimental Sec.7.



{\sl 6.  Experimental detections in materials:  }


  
\begin{figure}[tbhp]
\centering
\includegraphics[width=.7 \linewidth]{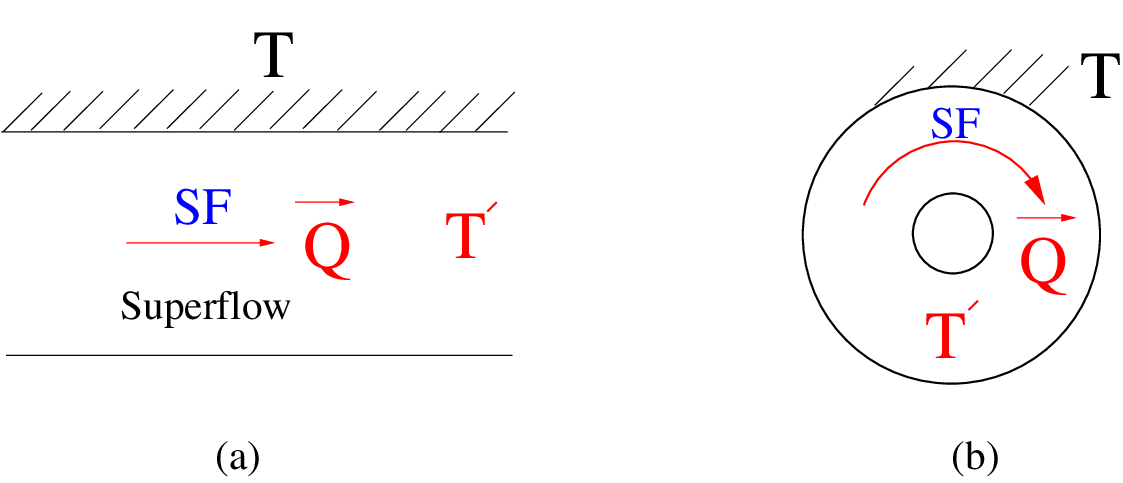}
\caption{ The temperature $T^{\prime} $ of the SF with a superflow $ \vec{Q} $ is slightly above
that  of its environment. 
The critical $ T^{\prime}_c $ in the co-moving frame will be slightly below that in the static sample.
(a) and (b) are stripe and annulus geometry respectively. The environment plays the role of the reservoir in a canonical 
or grand-canonical ensemble.   }
\label{movingSF}
\end{figure}   
  
  In terms of a moving sample, a running  superfluid (SF) Helium 4 may be a good candidate ( Fig.\ref{movingSF} ).
When $ T < T_c \sim 2.17 K $ below its critical temperature and a flowing speed $ v < v_c \sim 60 m/s $ below its critical velocity,
then $ (v/c)^2\sim 10^{-15} $.  Eq.\ref{Tlaw} shows that the bosonic excitation number in the lab frame satisfies:
\begin{equation}
   n^{\prime}( \omega^{\prime} ) d\omega^{\prime} = \frac{ D ( \omega^{\prime} ) d \omega^{\prime} }{e^{\frac{\hbar \omega^{\prime}}{k_B T^{\prime}}}-1 }
\label{bosonT}
\end{equation}
  where $  D ( \omega^{\prime} ) $ is the density of states (DOS) of the excitation spectrum and  $ T^{\prime} $ is the temperature:
  the former  $  \omega^{\prime} =\gamma(\omega-\vec{v} \cdot \vec{k}) $ is determined by the relative motion between the sample and the observer, it is odd under
  $ \vec{v} \to - \vec{v} $ ,   while the latter $ T^{\prime}=\gamma T  $ in Eq.\ref{Tlaw} is determined by the relative motion 
  between the sample and the reservoir, it is even under
  $ \vec{v} \to - \vec{v} $. Interestingly, $  \omega^{\prime}/T^{\prime}=(\omega-\vec{v} \cdot \vec{k})/T $ where the relativistic factor $ \gamma $ drops out.
  Of course, this simplification only holds for the case (A) where $ \gamma_{12}=\gamma_1 $. 
  So all the relativistic effects are encoded in the DOS in 
  Eq.\ref{bosonT} which can be mapped out by various by various neutron, light or X-ray scattering on a moving sample with a velocity $ \vec{v} $.



  Most directly, as shown in Eq.\ref{ZCo},  the body temperature of a running sample is independent of any inertial observer, so
  one may just attach a thermometer to the sample and read the temperature recorded by the thermometer.
  There are no Doppler shifts in the co-moving frame, but $ T_c \sim 2.17 K $ will be lowered in the co-moving frame according to Eq.\ref{Tclaw}.
  
In view of the recent experimental detection of the gravitational  red shift in a gravitational field by the atomic clocks  \cite{clock} to reach
the un-precedent accuracy $ 10^{-20} mm^{-1} $ ( see Eq.S15 ), we expect it is well within the current experimental precision measurement.
See also the precision measurement of the transverse Doppler shift in SM.
In fact, it is still much larger than the frequency shift  $ \Delta f/f \sim 10^{-21} $ caused by the gravitational wave which was detected by
the Laser interferometry.  For the Unruh effect  \cite{RMPUn}, taking the acceleration
$ a_p = 1 m/s^2 $ corresponds to $ T_U \sim 4 \times 10^{-21} K $ which is also much smaller than the temperature effects.
Unfortunately,  the temperature variations caused by several other factors such as the impurities or sample inhomogeneities
may be much larger than this order at $ (v/c)^2\sim 10^{-15} $. It remains challenging to remove these other factors.

{\sl 7.  Discussions.  }

   Our results can be  summarized as follows:
\begin{align}
  l/l_0 & = \gamma^{-1}_{12},~~ t/t_0 = \gamma_{12},  ~~~~~ m/m_0 =\gamma_{12}   \nonumber  \\
  T^{\prime}/T & = \gamma_1 > 1, ~~~~~T^{\prime}_c/T_c = \gamma^{-1}_1 < 1
\label{lists}
\end{align}
 where we also listed the well known results for the length, time and mass for a comparison and completeness.
 The law for the critical $ T_c $ only holds in the co-moving frame of a canonical ensemble.

 From a fundamental physics standpoint, space, time, and mass constitute the three most elemental elements, simultaneously establishing the three fundamental units: Centimeter (C), Second (S), and Gram (G). These units inherently rely on the relative motion between the sample and the observer, irrespective of the absence of a reservoir at $ T=0 $ or its presence at $ T > 0 $. The law of temperature transformation is contingent solely upon the relative motion between the sample and the reservoir, unaffected by the observer. This universality holds true, independent of microscopic details. Only in the case (A) depicted in Fig. 1 do the two relative motions coincide, resulting in 
 $ \gamma_{12}=\gamma_1 $. For any equilibrium system, the temperature can be treated as the imaginary time, elevating its significance to a status equivalent to that of time, as outlined in Eq. \ref{lists}. Notably, the temperature introduces a wholly new unit: Kelvin (K).
While we focused on the micro-canonical and canonical ensembles, our future endeavors will delve into studying how the chemical potential changes in a grand-canonical ensemble within a moving sample. It may also be interesting to extend to a non-equilibrium steady state 
such as the turbulence \cite{log} where the temperature may also be well defined. 



I  thank Fadi Sun and  C. L. Chen for very helpful discussions.
In the time order of my learning, I  thank  Yuhan Ma,  Jiangxin Lu, Mingzhe Li,
Jiangxin Lu, Hong Lv, Daulong Gao, Qun Wang,  Li You, Zhiyuan Sun, Weilan Guo,  Wenyu Wang and Haitao Quan for many helpful discussions.
I have always been indebted  to Guowu Meng and Jun Nie for long time consistent inspiring and stimulating  discussions.

{\bf Appendices }

{\bf  Difference between the high energy experiments and condensed matter experiments:  }
 In the high energy experiments such as LHC, RHIC or neutron stars \cite{ThermalQFT}, the effects in Eq.\ref{Tlaw} is  evident.
 Unfortunately, due to the large apparatus,  it is difficult to move the sample only without moving the reservoir, so
 it belongs to the case (B) in Fig.1. In fact, as shown in the Fig.3b, it becomes equivalent to the micro-canonical ensemble.
 Indeed, in the head-on heavy ion collisions, the temperature $ T \sim 10^{12} $ K can not be directly measured,  
 but can be deduced from the transverse momentum distributions of the scattered quarks
 in the center of momentum (COMM) frame. Their numbers  $ N \sim 10^{3} $  is much smaller than
 the sample in a condensed matter system. It can be described by a micro-canonical ensemble where the $ T $ is
 generated by self-collisions among these small number of quarks trapped around the zero COMM regime.
 On the other hand, for the putative electro-weak symmetry breaking transitions  at $ T \sim 10^{16} $ K in an early hot universe, it is not known if
 it can be described by an micro-canonical ensemble.

 However, in the condensed matter experiments,
 due to the small sample, the experimental set-up is automatically in the case (A) in Fig.1 and Fig.3a  where the small sample is moving
 relative to the large reservoir. Unfortunately,  the effects in Eq.\ref{Tlaw} are small due to $ v/c \ll 1 $.
 Then one must resort to the non-relativistic effects in a running superfluid. 
 In the main text, we present our experimental detections for the most common case (A), then the co-moving case
 which is a special case of the generic case of (A) +(B). The generic case can be similarly discussed.
 
  The transformation law of the thermodynamic potential   $ F= -\ln Z /\beta= E-TS   $  and all the physical quantities
 such as the entropy $ S $, the Pressure $ P $   automatically follow from $ dF=-S dT- PdV  $:
\begin{align}
 S_{AB}&= S^{\prime}[ \gamma^{-1}_1 \beta,   \gamma^{-1}_{12} V ],    \nonumber  \\
 P_{AB}&= P^{\prime}[ \gamma^{-1}_1 \beta,  \gamma^{-1}_{12} V ]
\label{ZpZSP}
\end{align}

 As stressed below Eq.\ref{ZpZVA},  the transformation laws of $ (T, V) $ are universal independent of if the systems are LI or not, so they also apply to
 non-relativistic systems such as  materials or cold atom systems.
 However, how do their corresponding 2 conjugate variables $ S, P $  transform are not universal, depend on the specific systems, 
 also on both $ v_1 $ and $ v_{12} $.

{\bf  The dynamics during the transit process:  }
  The static sample and the moving sample in Fig.M2 are the initial and final equilibrium state.
  They must be connected by a transit process.   In this subsection, we present the case (A) first, then the case (B). 
  They can be easily generalized to the other cases.
  
\begin{figure}[tbhp]
\centering
\includegraphics[width=.95 \linewidth]{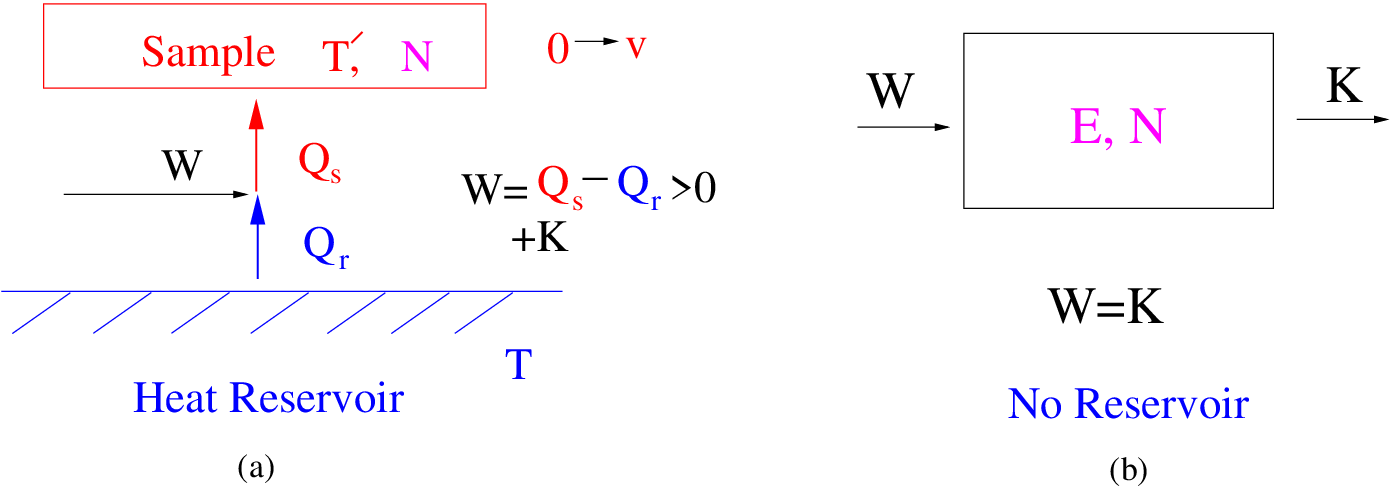}
\caption{ (a)  The transit process corresponding to (A) in Fig.M1 with a canonical ensemble.
The number $ N $ is fixed. During the transit process to accelerate the sample from rest to $ v $, its kinetic energy increase is $ K $,
its temperature increases from $ T $ to $ T^{\prime} $. The energy conservation leads to
Eq.\ref{W} . It applies to the case (A) in Fig.1.
(b) The transit process corresponding to (B) in Fig.M1 which is equivalent to a micro-canonical ensemble.
No external reservoir.
The internal energy $ E $ and the number of particles $ N $ are fixed.
Then the external mechanical work done on the sample $ W $ is just the increase of its kinetic energy $ W =  K $. It applies to the case (B) in Fig.1. }
\label{heatqmicro}
\end{figure}

\begin{figure}[tbhp]
\centering
\includegraphics[width=0.5 \linewidth]{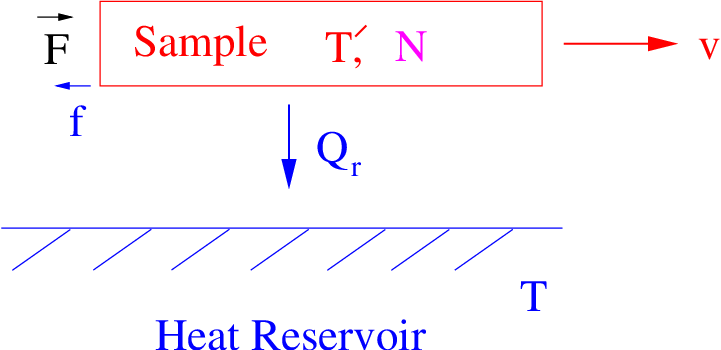}
\hspace{0.2cm}
\includegraphics[width=.35 \linewidth]{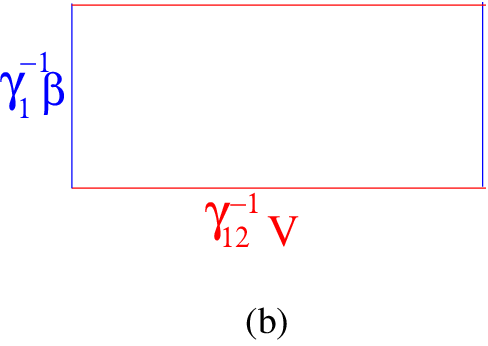}
\caption{ (a) The equilibrium state of the case (A). 
The relative motion between the system and the reservoir leads to a very small friction force $ \vec{f} $  between the sample and heat  reservoir.
One may need to apply an infinitesimal external force $ \vec{F}= - \vec{f} $ to keep the sample moving at a constant velocity $ \vec{v} $,
therefore transfers a small heat and entropy into the reservoir. (b)  The unified  geometric interpretation of the transformation laws of the temperature and volume:
both can be viewed as the scale of $ \gamma^{-1}_1 < 1 $ and 
$ \gamma^{-1}_{12} < 1 $  in the finite size along the imaginary time direction and the real space direction  respectively.  }
\label{steady}
\end{figure}

  One can  compute the entropy production in the sample during the transit process:
\begin{align}
 \Delta S_s= \int^{T^{\prime}}_{T}  C_v \frac{dT}{T}
\label{heat}
\end{align}
  For a solid,  $ C_v \sim T^3 $ at low $ T $ ( not too low such that the electronic contribution can be ignored ),
  then $ \Delta S_s \sim T^3 ( \gamma^3 -1 ) > 0 $,
  $ C_v \sim const. $ at a high $ T $, then $ \Delta S_s \sim  \log \gamma > 0 $.
  The corresponding heat transferred to the sample $ Q_s =\int^{T^{\prime}}_{T}  C_v dT > 0 $ is
  $  Q_s \sim T^4 ( \gamma^4 -1 )  $ and  $ Q_s \sim T ( \gamma-1 ) $ at a low and high $ T $ respectively.
  It comes from the external work and the  reservoir $ Q_s= W-K + Q_r $  ( Fig.\ref{heatqmicro}a ). 
  Note that the reservoir does not do any mechanical work on the sample during the process.
  
  The external mechanical work done on the sample
\begin{align}
  W =Q_s- Q_r + K
\label{W}
\end{align}
  where $ Q_s- Q_r > 0 $ is the difference between the heat $ Q_s $ transferred to the sample and that $ Q_s $ extracted from the reservoir.
  So the transit process may behave like a refrigerator.

  The second law requires:
\begin{align}
 \Delta S= \Delta S_s-\frac{Q_r}{T} \geq 0
\label{third}
\end{align}
  where $ W $ and $ Q_r $ depend on the transit process in Fig.\ref{heatqmicro}a. It dictates the in-equality
\begin{align}
  W-K \geq Q_s-T \Delta S_s 
\label{third2}
\end{align}
   which reduces to $ W-K \geq T ( \gamma-1 -\log \gamma ) > 0  $ at low $ T $ and
   $ W-K \geq T ^4 \gamma^3( \gamma-1 ) > 0  $ at high $ T $.
   The equality holds only for an adiabatic process which keeps the total entropy invariant.
   
  Now we can look at the steady state in the Fig.1A.  The relative motion between the system and the reservoir leads to a very small friction force 
   between the sample and heat  reservoir $
  \vec{f} \sim - \epsilon T \vec{v}  \to 0 $
  where $ \epsilon \to 0  $ stands for the very small interaction between the system and the reservoir.
  Of course, it vanishes in a static sample and at zero temperature. So to keep its motion, one may apply
  a constant external force $ \vec{F}=  - \vec{f} \sim \epsilon T \vec{v} $, so the work done by the external force 
  is $ \vec{F} \cdot \vec{v}  \sim \epsilon T v^2  $ which transfers heat $ Q_r \sim \epsilon T v^2 $  into the reservoir, therefore releases the entropy
  $ S_r \sim  \epsilon v^2 $ into the reservoir ( Fig.\ref{steady} ). As argued below, incorporating this very small effect will not change any results
  without it. Of course, the running superfluid presented in the Sec.7 does not even have such an infinitesimal friction.

  In fact, the total energy of the system plus the reservoir is $ H_t=H_s + H_r + \epsilon H_{sr} $ where $ \epsilon \ll 1 $.
  If bringing the system and the reservoir at Temperature $ T $ together, it takes about the relaxation time  $ t_R \sim 1/\epsilon^2 $ for 
  the system to reach the equilibrium with the reservoir \cite{Pbook}. However, after reaching the equilibrium, one may just set 
  $ \epsilon \to 0  $ and simply write $ H_t=H_s + H_r  $ which leads to the probability distribution and partition function in a canonical ensemble \cite{kardarbook}.
  The subtlety here is that on the one hand, one clearly needs the interaction $ \epsilon $ term to exchange heat
  between the system and  the reservoir  and establish the thermal equilibrium between the two, 
  on the other hand, to describe an equilibrium system, one needs to send  $ \epsilon \to 0  $.
  Otherwise $ \epsilon \sim 1 $ will put the system into an open dissipative system where the temperature is not even defined.


  Now we look at the transition process for the case (B) in Fig.M1 and Fig.\ref{heatqmicro}b:
  during the transit process to accelerate the sample from rest to $ v $, its kinetic energy increase is $ K $.
  Because the sample and the reservoir  are tied together, there is  no heat transfer between
  the two,  so it is equivalent to a micro-canonical ensemble with a fixed
  internal energy $ E $ and the number of particles $ N $.
  Then the external mechanical work done on the sample is just the increase of its kinetic energy $ W =  K $,
  so its temperature  remains the same.
  

  In short, despite the two ensembles:  the canonical and micro-canonical ensembles make no difference in the
  thermodynamic limit in the static sample, they do make differences when the sample is moving.
  If incorporating the kinetic energy into the thermodynamic equation, one can write the total energy:
  $ dE_t= T dS-P dV+ \vec{v} \cdot d \vec{P} $. For the initial and final state in  Fig.\ref{heatqmicro}, the last term can be dropped.
  But as said above it is useful in the transit process connecting the initial to the final state.
  As to be discussed in Sec.7, the canonical and micro-canonical ensembles  also make differences in the black hole thermodynamics and AdS/CFT correspondence.   
  More physical interpretations will be given in the appendix D.

{\bf Using the Eigenstate Thermalization hypothesis on the micro-canonical ensemble :  }

 In modern quantum statistical mechanics, there is a renewed interest on Eigenstate Thermalization hypothesis
(ETH) in many body interacting systems. It states that
 for any excited ( also called bulk ) eigen-state $ |\psi \rangle $ with an eigen-energy $ E $ which is above the ground state energy by a finite
 amount in the thermodynamic limit: $ \lim_{V \rightarrow \infty} \frac{E-E_0}{V} \neq 0 $, one may define
 a temperature $ \beta=1/k_B T $ corresponding to the state: 
 \begin{align}
  \langle \psi | H | \psi \rangle = Tr H e^{-\beta H }/Z 
\label{ethH}
\end{align}
 where $ Z= Tr e^{-\beta H } $ is the  partition function in a canonical ensemble.
 It taking $ |\psi \rangle $ to be the ground state $ |\psi_0 \rangle $ with an eigen-energy $ E_0 $, then the corresponding $ T=0 $.
 
  The ETH implies that for any local operator $ O $:
\begin{equation}
  \langle \psi | O | \psi \rangle = Tr O e^{-\beta H }/Z
\label{eth}
\end{equation}
 The entanglement entropy of the excited  state $ | \psi \rangle $ satisfies the
 volume law, while that of  the ground state or all the low energy states above the ground state  satisfy the more common area law.
 Despite there is no rigorous mathematical proof from the probability theory yet on the ETH, there are convincing numerical evidences 
 that it holds in any quantum chaotic systems of which the SYK model \cite{SY,kittalk,syk2,SYKETH}  is just an simple example.
 
 This mapping between the many-body eigen-state $ |\psi \rangle $ in a micro-canonical ensemble and the temperature of a canonical  ensemble
 maybe used to define the temperature of an isolated self-interacting many body system.
 Then it may be applied to show that the temperature remains the same Eq.\ref{TT0}  in  the case (A) in Fig.1 or equivalently the Fig.4b. 
 Indeed, any eigen-state $ |\psi \rangle $  may suffer at most a unitary transformation under the LT, so it will not change the LHS of
 Eq. \ref{ethH}, which implies the temperature defined on the RHS remains the same.
 It remains unknown how to push the ETH for the case (B) to the much more interesting case (A).








\end{document}